\documentclass[aps,prb,reprint,amsmath,amssymb,nosuperscriptaddress]{revtex4-1}
\usepackage{amsmath}
\usepackage{amssymb}
\usepackage{bm}
\usepackage{color}
\usepackage[colorlinks]{hyperref}
\usepackage{mathrsfs}

\let\de=\partial

\newcommand\dd{\mathrm{d}}
\newcommand\La{\mathscr{L}}
\newcommand\Ha{\mathscr{H}}

\newcommand\Na{\mathscr{N}}

\newcommand\vek[1]{\bm{#1}}
\newcommand\imag{\mathrm{i}}
\newcommand\he[1]{#1^{\dagger}}% hermitian conjugate
\newcommand\gr[1]{\mathrm{#1}}
\definecolor{myred}{rgb}{1,0,0}

\begin{document}
\title{Remarks on relativistic scalar models with chemical potential}

\author{Tom\'{a}\v{s} Brauner}
\email{tomas.brauner@uis.no}
\affiliation{Department of Mathematics and Physics, University of Stavanger, 4036 Stavanger, Norway}

\begin{abstract}
We discuss selected aspects of classical relativistic scalar field theories with nonzero chemical potential. First, we offer a review of classical field theory at nonzero density within the Lagrangian formalism. The aspects covered include the question of equivalence of descriptions of finite-density states using a chemical potential or time-dependent field configurations, the choice of Hamiltonian whose minimization yields the finite-density equilibrium state, and the issue of breaking of Lorentz invariance. Second, we demonstrate how the low-energy effective field theory for Nambu-Goldstone (NG) modes arising from the spontaneous breakdown of global internal symmetries can be worked out explicitly by integrating out the heavy (Higgs) fields. This makes it possible to analyze the spectrum of NG modes and their interactions without having to deal with mixing of NG and Higgs fields, ubiquitous in the linear-sigma-model description of spontaneous symmetry breaking.
\end{abstract}

\maketitle

%%%%%%%%%%%%%%%%%%%%%%%%%%%%%%%%%%%%%%%%%%%%%%%%%%%%%%%%%%%%%%%%%%%

\section{Introduction}
\label{sec:intro}

Scalar field theory, both classical and quantum, constitutes a paradigm for the study of ordered phases of matter.\cite{Ginzburg1950a,*Goldstone1961a,*Goldstone1962a,*Higgs1964a} In the early days, heuristic scalar models offered a welcome, practically convenient way to explore the then novel phenomenon of spontaneous symmetry breaking in as simple a setting as possible. From the modern perspective of effective field theory (EFT), scalar field theory is an inevitable part of our understanding of spontaneously broken internal symmetries. Indeed, it accounts for the universal critical properties of continuous phase transitions, as well as provides a model-independent EFT description of the long-distance physics of an ordered phase away from the phase transition.

In addition to the model-independent EFT framework, specific scalar models continue to provide an invaluable playground for investigation of novel phenomena where spontaneous symmetry breaking plays a key role. This applies especially to the physics of quantum many-body systems. In vacuum, the structure of both the spectrum of particles and their interactions is severely restricted by Lorentz invariance. A minimal modification required to describe actual matter is to add a set of chemical potentials for the conserved charges that acquire a nonzero value in the  statistical equilibrium state of the system.

Indeed, this is how the intricacies of the spectrum of Nambu-Goldstone (NG) bosons arising from spontaneous breakdown of a continuous global internal symmetry in a quantum many-body system\cite{Nielsen1976a,*Nambu2004a,Leutwyler1994a} were first understood.\cite{Miransky2002a,*Schaefer2001a,*Brauner2005a,Brauner2010a} Later on, this understanding was supplemented by a model-independent EFT description of the NG bosons,\cite{Watanabe2012b,*Watanabe2014a} valid for relativistic and nonrelativistic systems alike.

In this note, we address several issues pertinent to the description of dense matter using (classical) relativistic scalar field theory. Section~\ref{sec:density} is conceptual and the discussion is therefore limited to the simple class of models with a single complex scalar field and $\gr{U(1)}$ internal symmetry, corresponding to particle number. We start from scratch by seeking field configurations that carry nonzero particle number density, and remind the reader how the corresponding statistical equilibrium state can be equivalently described using a chemical potential. We then explain how the setup is affected by explicit breaking of the $\gr{U(1)}$ symmetry. Having a clear picture of the relation between the different descriptions of finite-density matter, and their limitations, also sheds light on the subtle manifestation of Lorentz invariance in presence of a chemical potential. The material contained in this section is not new. It seems, however, that a unified discussion of the above issues is not available elsewhere.

In Sec.~\ref{sec:EFT}, we then discuss the low-energy EFT for NG bosons of the spontaneously broken global continuous symmetry. In order to make the results of practical utility, we cover a larger class of models with non-Abelian internal symmetry here. The routine for analyzing spontaneous symmetry breaking in a scalar field theory at the classical (tree) level is: minimize the Hamiltonian to find the ground state, expand the Lagrangian in fluctuations of fields around the ground state, identify the spectrum of excitations from the bilinear part of the Lagrangian and the interactions from the higher-order terms. In presence of a chemical potential, one as a rule has to deal with field mixing in the kinetic term; the mass matrix alone is not sufficient to extract the dispersion relations of the various excitations. If one is only interested in the physics of NG bosons, then the mixing issues can be avoided by integrating out the Higgs mode(s). We demonstrate how this can be done explicitly by an iterative solution of the equation of motion for the Higgs mode(s).

%%%%%%%%%%%%%%%%%%%%%%%%%%%%%%%%%%%%%%%%%%%%%%%%%%%%%%%%%%%%%%%%%%%

\section{Finite-density states}
\label{sec:density}

Let us consider the following class of Lagrangian densities for a single complex scalar field $\phi$,
\begin{equation}
\La=\de_\mu\phi^*\de^\mu\phi-V(\phi^*\phi)+(j^*\phi+j\phi^*);
\label{Lag}
\end{equation}
the specific form of the classical potential $V$ is not important. For vanishing source, $j=0$, the Lagrangian has an exact global $\gr{U(1)}$ symmetry under the change of phase of $\phi$. For the time being, we will indeed assume that $j=0$, and return to the effects of nonzero $j$ later on.

The Noether current associated with the $\gr{U(1)}$ symmetry of the Lagrangian~\eqref{Lag} is
\begin{equation}
J^\mu=\imag(\phi^*\de^\mu\phi-\de^\mu\phi^*\phi).
\label{current}
\end{equation}
Upon using the classical Euler-Lagrange equation for $\phi$, the current is seen to satisfy the conservation law
\begin{equation}
\de_\mu J^\mu=-\imag(j^*\phi-j\phi^*).
\end{equation}
Nonzero density of the $\gr{U(1)}$ charge (dubbed as ``particle number'') requires that the field $\phi$ depends nontrivially on time. The correspondence between particle number and time dependence of the field can be specified more concretely if one restricts to special many-body states. In statistical mechanics, many-particle systems in equilibrium are described using the grand canonical ensemble, which in the limit of zero temperature corresponds to the state of lowest energy at fixed (average) particle number. How to implement this definition in classical field theory is detailed in the next two subsections.

%%%%%%%%%%%%%%%%%%%%%%%%%%%%%%%%%%%%%%%%%%%%%%%%%%%%%%%%%%%%%%%%%%%

\subsection{Canonical approach}

The Lagrangian~\eqref{Lag} describes completely the dynamics of the classical field $\phi$ given proper initial conditions: the values of the field, $\phi(t_i,\vek x)$, and its temporal derivative, $\dot\phi(t_i,\vek x)$, at the chosen initial time $t=t_i$ and everywhere in space. The corresponding canonical Hamiltonian density takes the form
\begin{equation}
\Ha=\Pi^*\Pi+\vek\nabla\phi^*\cdot\vek\nabla\phi+V(\phi^*\phi)-(j^*\phi+j\phi^*),
\label{Ham}
\end{equation}
where $\Pi=\dot\phi^*$ is the canonical momentum variable conjugate to $\phi$. In the Hamilton formalism, the time evolution is fully specified by the initial values $\phi(t_i,\vek x)$ and $\Pi(t_i,\vek x)$.

A state minimizing the total energy subject to the constraint of a fixed particle number can be found as usual using the method of Lagrange multipliers. This amounts to replacing the Hamiltonian~\eqref{Ham} with
\begin{equation}
\Ha_\mu\equiv\Ha-\mu\Na,\qquad
\Na\equiv J^0=\imag(\Pi^*\phi^*-\Pi\phi).
\label{muHam}
\end{equation}
It is easy to see what the field configuration minimizing (the spatial integral of) $\Ha_\mu$ is. By rewriting $\Ha_\mu$ as
\begin{equation}
\begin{split}
\Ha_\mu={}&|\Pi-\imag\mu\phi^*|^2+\vek\nabla\phi^*\cdot\vek\nabla\phi\\
&+V_\text{eff}(\phi^*\phi)-(j^*\phi+j\phi^*),
\end{split}
\label{muHamaux}
\end{equation}
where $V_\text{eff}(\phi^*\phi)\equiv V(\phi^*\phi)-\mu^2\phi^*\phi$, we find that for $j=0$, the two lines of Eq.~\eqref{muHamaux} can be minimized separately. The ground state is then a spatially uniform field configuration $\phi(t)$ whose magnitude minimizes the effective potential $V_\text{eff}(\phi^*\phi)$.\footnote{The Hamiltonian $\Ha_\mu$ can be minimized in the same way even for nonzero $j$, but then the time evolution implied by the relation $\Pi=\dot\phi^*$ takes the field outside the thus found ground state manifold.} The time dependence is determined by the very first term in $\Ha_\mu$, minimized by $\Pi=\imag\mu\phi^*$, which in turn implies
\begin{equation}
\phi(t)=\phi(0)e^{-\imag\mu t}.
\label{tphase}
\end{equation}
For any potential $V$, bounded from below and asymptotically growing faster than $\phi^*\phi$, the effective potential $V_\text{eff}(\phi^*\phi)$ will acquire its absolute minimum at some nonzero value of $\phi$ for sufficiently large chemical potential. This is relativistic Bose-Einstein condensation.

While the grand canonical Hamiltonian~\eqref{muHam} serves to define a specific many-body \emph{state}, the time evolution is still governed by the Hamiltonian~\eqref{Ham}, hereafter referred to as ``microscopic.'' This observation formed the basis for a prediction of NG-like modes, protected by symmetry but with a nonzero gap fixed by symmetry and the chemical potential alone.\cite{Nicolis2013a} This prediction was subsequently extended from relativistic systems with chemical potential to all local quantum field theories, relativistic or not.\cite{Watanabe2013b}

Motivated by the time dependence of the Bose-Einstein condensate~\eqref{tphase}, it is common in statistical mechanics to ``redefine'' time evolution by stripping off the factor $e^{-\imag\mu t}$. What such a ``redefinition'' of time evolution means can be understood as follows. Start by changing the configuration space variable $\phi$ to $\varphi$ through
\begin{equation}
\phi(t,\vek x)\equiv e^{-\imag\mu t}\varphi(t,\vek x).
\label{subs}
\end{equation}
This is just a substitution of variables that has no physical content. It has, however, the advantage that the condensate~\eqref{tphase} is represented by a uniform, time-dependent $\varphi$. The resulting grand canonical Lagrangian, stemming from the microscopic Lagrangian~\eqref{Lag}, reads
\begin{equation}
\begin{split}
\La_\mu={}&D_0\varphi^*D_0\varphi-\vek\nabla\varphi^*\cdot\vek\nabla\varphi-V(\varphi^*\varphi)\\
&+(e^{-\imag\mu t}j^*\varphi+e^{\imag\mu t}j\varphi^*),
\end{split}
\label{muLag}
\end{equation}
where $D_0\varphi\equiv\dot\varphi-\imag\mu\varphi$. The substitution of variables~\eqref{subs} can be lifted to the phase space, where it has to be augmented with $\Pi(t,\vek x)\equiv e^{\imag\mu t}\pi(t,\vek x)$, so that the resulting change from $\phi$, $\Pi$ to $\varphi$, $\pi$ is a canonical transformation.

In terms of the new canonical variables, the Hamilton equation of motion for $\Pi$ can be rewritten as
\begin{equation}
-\frac{\delta\Ha}{\delta\phi}=\dot\Pi=e^{\imag\mu t}(\dot\pi+\imag\mu\pi),
\end{equation}
which implies that
\begin{equation}
\dot\pi=-\frac{\delta\Ha}{\delta\varphi}-\imag\mu\pi=-\frac{\delta\Ha}{\delta\varphi}+\mu\frac{\delta\Na}{\delta\varphi}.
\label{subspi}
\end{equation}
A similar result is found for $\dot\varphi$. These equations of motion are fully equivalent to the original ones, stemming from the Lagrangian~\eqref{Lag}, regardless of the source~$j$, that is, whether the $\gr{U(1)}$ symmetry is exact or not. One should however be careful here: the effective Hamiltonian on the right-hand side of Eq.~\eqref{subspi} was obtained by a substitution of variables, hence it is given by $\tilde\Ha_\mu(\varphi,\pi)\equiv\tilde\Ha(\varphi,\pi)-\mu\tilde\Na(\varphi,\pi)$, where
\begin{equation}
\begin{split}
\tilde\Ha(\varphi,\pi)&\equiv\Ha(\phi(\varphi),\Pi(\pi)),\\
\tilde\Na(\varphi,\pi)&\equiv\Na(\phi(\varphi),\Pi(\pi)).
\end{split}
\end{equation}
This makes it clear that apart from defining the statistical equilibrium, the grand canonical Hamiltonian~\eqref{muHam} also gives an equivalent description of the time evolution of the fields \emph{provided} that $\tilde\Ha(\varphi,\pi)=\Ha(\varphi,\pi)$ and $\tilde\Na(\varphi,\pi)=\Na(\varphi,\pi)$. The latter condition is equivalent to the requirement of exact $\gr{U(1)}$ symmetry.

Let us briefly summarize with a series of comments. First, the dynamics of the considered class of classical field theories is defined by default by the microscopic Lagrangian~\eqref{Lag} or the microscopic Hamiltonian~\eqref{Ham}. When $j=0$, that is when the theory~\eqref{Lag} possesses an exact $\gr{U(1)}$ symmetry, the dynamics can be equivalently described by stripping off a time-dependent phase as in Eq.~\eqref{subs} and shifting the Hamiltonian according to Eq.~\eqref{muHam}. This observation per se does not depend on the assumption of statistical equilibrium. However, with this assumption, the constant $\mu$ can be interpreted as a chemical potential, defining certain stationary many-body state.

Second, when $j=0$, the chemical potential enters the grand canonical Lagrangian~\eqref{muLag} solely through the covariant derivative $D_0\varphi$, hence it acts as a constant background temporal gauge field. This is a very general feature of the Legendre transform connecting the Hamilton and Lagrange pictures, not limited to the class of Hamiltonians~\eqref{Ham} or to $\gr{U(1)}$ symmetries.\cite{Kapusta1981a}

Finally, when $j\neq0$, that is when the $\gr{U(1)}$ symmetry is explicitly broken, most of the arguments above break down. It is still true that the time evolution of the field $\phi$ is governed by the Lagrangian~\eqref{Lag} or the Hamiltonian~\eqref{Ham}. However, this time evolution can no longer be recovered by naively inserting the chemical potential in the Lagrangian through covariant derivatives acting on the phase-redefined field $\varphi$. This is, after all, already obvious from Eq.~\eqref{muLag}. Analogously, one cannot recover the correct time evolution by naively replacing the microscopic Hamiltonian~\eqref{Ham} with the grand canonical Hamiltonian~\eqref{muHam} in the Hamilton equations of motion. We can still formally use $\mu$ as a Lagrange multiplier to define the initial condition for time evolution through constrained minimization of the Hamiltonian. However, the thus obtained many-body state will in general not be a stationary state of the Hamiltonian~\eqref{Ham}.

In short, we conclude that the fundamental description of a many-body state is that in terms of a time-dependent field configuration, whose complex phase encodes the chemical potential. It is common to instead introduce the chemical potential as a constant background temporal gauge field in the Lagrangian, mostly due to the practical advantages of the Lagrange formalism over the Hamilton one. One must, however, keep in mind that the latter implementation of the chemical potential is derived from the former one, and rests on the assumption of \emph{exact} $\gr{U(1)}$ symmetry.

%%%%%%%%%%%%%%%%%%%%%%%%%%%%%%%%%%%%%%%%%%%%%%%%%%%%%%%%%%%%%%%%%%%

\subsection{Noether current approach}

The canonical approach is not always convenient, especially when the Lagrangian depends on higher derivatives of the fields. This is typical for low-energy EFTs of NG bosons. In such a case, it is advantageous to exploit the above-made observation that the chemical potential can be treated as a constant temporal background gauge field of the particle number $\gr{U(1)}$ symmetry. This implementation of the chemical potential within EFT can be justified provided that the underlying microscopic theory is a renormalizable field theory, which can be treated with the canonical approach of the previous subsection.\footnote{It is commonplace to assume that a given low-energy EFT can be ``UV-completed'' to a microscopic renormalizable field theory, although there are other logical possibilities. On the one hand, the microscopic theory need not be a local field theory. On the other hand, it is possible that there is no complete microscopic theory at all, and one instead finds an infinite tower of EFTs with successively broader range of validity.\cite{Castellani2002a}}

Formally, one replaces the chemical potential with a $\gr{U(1)}$ gauge field $A_\mu$ acting as a source for the particle number current. The generating functional $Z[A_\mu]$ of the microscopic theory then has $\gr{U(1)}$ gauge invariance that must be reproduced by the low-energy EFT.\cite{Leutwyler1994a,Leutwyler1994b,*Son2002a} The presence of an explicit symmetry-breaking term in the microscopic theory is not necessarily an obstacle. The exact $\gr{U(1)}$ symmetry of the Lagrangian~\eqref{Lag} can be rescued by promoting the parameter $j$ to a field that transforms simultaneously with $\phi$ and  in the same way as $\phi$. We then have a generating functional $Z[A_\mu,j]$ that still possesses exact $\gr{U(1)}$ gauge invariance.

The background field approach is extremely useful as it strongly constrains the dependence of the low-energy EFT on both the chemical potential and the parameter $j$ explicitly breaking the global $\gr{U(1)}$ symmetry. It comes, however, with subtleties. Namely, in order to identify a finite-density equilibrium state of the EFT, we still need a suitable Hamiltonian to be minimized. In the absence of a simple canonical structure due to the dependence of the effective Lagrangian on higher field derivatives, it is natural to obtain the Hamiltonian using Noether's theorem. This procedure is, however, known to suffer from ambiguities, see Ref.~\onlinecite{Brauner2020a} for a recent elementary account.

In addition, within the Lagrangian formalism, one is free to redefine a complex scalar field by Eq.~\eqref{subs}; this is equivalent to a particular $\gr{U(1)}$ gauge transformation. In this way, one can completely remove the chemical potential from the Lagrangian or give it any nonzero value. This reflects the fact that since both $\Ha$ and $\Na$ are conserved charge densities, so is any linear combination of theirs. The freedom to  redefine the field variable of the EFT seems to prevent us from identifying a unique grand canonical Hamiltonian.

Here the symmetry-breaking parameter $j$, which was previously a nuisance, proves actually helpful. At this stage, one has to recognize that the gauge invariance of the generating functional $Z[A_\mu,j]$ is not physical; it is merely a useful tool to constrain the EFT. The true physical symmetry consists of transformations that leave the background fields $A_\mu$ and $j$ fixed. At $j=0$, the theory has separate global symmetries under time translations and under $\gr{U(1)}$ phase transformations. Any $j\neq0$ breaks the two independent symmetries to a particular ``diagonal'' combination of theirs.

In presence of a chemical potential $\mu$ and a constant, time-independent $j$, the exact symmetry is the usual time translation and the application of Noether's theorem leads to the usual grand canonical Hamiltonian. Upon redefining the field $\phi$ by Eq.~\eqref{subs}, the chemical potential disappears from the kinetic term in the Lagrangian~\eqref{Lag} but the source term becomes $e^{-\imag\mu t}j^*\varphi+e^{\imag\mu t}j\varphi^*$. Time translations are now explicitly broken and so is the $\gr{U(1)}$ symmetry. The theory nevertheless remains invariant under the combined transformation
\begin{equation}
t'=t+\theta,\qquad
\varphi'(t',\vek x)=\varphi(t,\vek x)e^{\imag\mu\theta}.
\label{mixtransfo}
\end{equation}
Noether's theorem then predicts the existence of a unique conserved charge that can be identified with the Hamiltonian of the system. It is easy to check that the symmetry~\eqref{mixtransfo} leads to the \emph{same} grand canonical Hamiltonian as the pure time translation in the case of constant, time-independent $j$; it is merely expressed in terms of the new field variable $\varphi$. This is hardly surprising; \emph{no} field redefinition can change the conserved charges of the theory.

The interplay of Noether's theorem and redefinitions of the field that helped us to pin down the unique grand canonical Hamiltonian also sheds light on the question of breakdown of Lorentz invariance in presence of a chemical potential. It used to be a common lore to claim that adding the chemical potential to the Lagrangian via covariant derivatives as in Eq.~\eqref{muLag} breaks Lorentz invariance explicitly. This view has been contested for some years by Nicolis and collaborators\cite{Nicolis2013a,Nicolis2013b,*Nicolis2015a} who argue that the breakdown of Lorentz invariance is spontaneous.

Indeed, in spite of appearances, treating finite-density states by adding a chemical potential to the Lagrangian has all attributes of spontaneous, not explicit, symmetry breaking. First, the chemical potential is a property of a particular many-body \emph{state}, not of the dynamics, which is still governed by the microscopic, manifestly Lorentz-invariant, Lagrangian~\eqref{Lag}. Second, even in presence of a chemical potential, Lorentz invariance still gives a set of \emph{exact} constraints on the Green's functions of the theory, which can be implemented conveniently used the background gauge invariance of the generating functional.\cite{Leutwyler1994a,Leutwyler1994b,*Son2002a} Finally, since the chemical potential can be removed from the Lagrangian by a mere field redefinition, it follows at once that the symmetries of the microscopic Lagrangian~\eqref{Lag} and the grand canonical Lagrangian~\eqref{muLag} must be in one-to-one correspondence. This implies that the grand canonical Lagrangian~\eqref{muLag} still has \emph{exact} Lorentz invariance. Lorentz boosts are just realized nonlinearly, in accord with the fact that they are spontaneously broken by the presence of the dense medium.

Before closing the general discussion of finite-density states in classical field theory, let us comment on another possible way to obtain the Hamiltonian from given action. Analogously to replacing the chemical potential $\mu$ and the symmetry-breaking parameter $j$ by background fields $A_\mu(x)$ and $j(x)$, one can similarly couple the theory to a background spacetime geometry, represented by the metric $g_{\mu\nu}(x)$. This gives a generally-covariant classical action and a gauge-invariant generating functional $Z[A_\mu,j,g_{\mu\nu}]$. A standard procedure to obtain a symmetric energy-momentum tensor is to take the variation of the action with respect to the metric. It is very instructive to check for oneself that for relativistic field theories with a chemical potential, this procedure leads to a \emph{wrong} Hamiltonian, differing from the grand canonical Hamiltonian~\eqref{muHamaux}. This is related to the above-mentioned ambiguity in the definition of the energy-momentum tensor using Noether's theorem. Further details are provided in Appendix~\ref{app:metric} for an interested reader.

%%%%%%%%%%%%%%%%%%%%%%%%%%%%%%%%%%%%%%%%%%%%%%%%%%%%%%%%%%%%%%%%%%%

\section{Low-energy EFT}
\label{sec:EFT}

In this section we consider a more general class of linear sigma models, including a set of $N$ complex scalar fields, collectively denoted as $\phi$. Just like for the $\gr{U(1)}$-invariant models of a single complex scalar discussed in the previous section, a nontrivial ground state appears when a sufficiently large chemical potential is introduced. We show how to obtain the low-energy EFT for the resulting NG modes by integrating out the Higgs mode, which on the classical level means solving its equation of motion.

In order that our derivation of the low-energy EFT can be as explicit as possible, we start with the following class of Lagrangians,
\begin{equation}
\La=D_\mu\he\phi D^\mu\phi-m^2\he\phi\phi-\lambda(\he\phi\phi)^2,
\label{LagN}
\end{equation}
where $D_\mu\phi\equiv(\de_\mu-\imag\delta_{\mu0}\mu)\phi$. When $\mu=0$, the Lagrangian has a global internal $\gr{SO}(2N)$ symmetry under rotations of all the $2N$ real degrees of freedom contained in $\phi$. The chemical potential $\mu$ breaks this down to the $\gr{SU}(N)\times\gr{U(1)}$ subgroup.\cite{Andersen2007a}

When $\mu^2-m^2>0$, the corresponding Hamiltonian is minimized by a nonzero $\langle\phi\rangle=v\phi_0$, where\footnote{For positive $m^2$, a chemical potential $\mu>m$ is required; this is relativistic Bose-Einstein condensation. For negative $m^2$, the ground state is nontrivial even for zero chemical potential; this is the usual Higgs model.}
\begin{equation}
v=\sqrt{\frac{\mu^2-m^2}{2\lambda}},
\end{equation}
and $\phi_0$ is normalized so that $\he\phi_0\phi_0=1$ but otherwise can assume an arbitrary orientation in the field space. The field $\phi$ is then parametrized in terms of fluctuations around this ground state,
\begin{equation}
\phi=(v+H)U(\pi)\phi_0,
\end{equation}
where $H$ is the amplitude (Higgs) mode and the unitary matrix $U(\pi)$ contains the NG fields, denoted as $\pi$. Inserting this parametrization in the Lagrangian~\eqref{LagN}, it becomes, up to an additive constant corresponding to the vacuum energy,
\begin{equation}
\La=(\de_\mu H)^2+(v+H)^2\Xi-4\lambda v^2H^2-4\lambda vH^3-\lambda H^4,
\label{LagN2}
\end{equation}
where
\begin{equation}
\Xi\equiv\he\phi_0\de_\mu\he U\de^\mu U\phi_0+2\imag\mu\he\phi_0\he U\de_0U\phi_0.
\label{Xidef}
\end{equation}

We now integrate out the Higgs field $H$ in a series of steps, taking advantage of the fact that the Lagrangian~\eqref{LagN2} depends on the NG fields solely through the expression $\Xi$. First, we deduce the equation of motion for the Higgs mode,
\begin{equation}
-\Box H+(v+H)\Xi-4\lambda v^2H-6\lambda vH^2-2\lambda H^3=0,
\label{EOM}
\end{equation}
which can be formally solved as
\begin{equation}
H=(4\lambda v^2-\Xi+\Box+6\lambda vH+2\lambda H^2)^{-1}v\Xi.
\label{higgs}
\end{equation}
Now $\Xi$ is small at low energies thanks to the fact that it contains derivatives. Consequently, $H$ is small as well and the denominator in Eq.~\eqref{higgs} is dominated by the first term, $4\lambda v^2$. In principle, we could now proceed by expanding Eq.~\eqref{higgs} iteratively and inserting the result back to the Lagrangian~\eqref{LagN2}. There is however another, more convenient way to manipulate the Lagrangian using the equation of motion~\eqref{EOM}. 

The sum of the last three terms in Eq.~\eqref{EOM} can be factorized and the equation of motion thus rewritten as
\begin{equation}
\Xi-2\lambda H(2v+H)=\frac{\Box H}{v+H}.
\label{EOM2}
\end{equation}
Multiplying Eq.~\eqref{EOM} by $H$ and subtracting it from the Lagrangian~\eqref{LagN2}, the latter becomes
\begin{equation}
\La=v(v+H)\Xi+2\lambda vH^3+\lambda H^4.
\end{equation}
Further manipulation using Eq.~\eqref{EOM2} then leads to
\begin{equation}
\La=v^2\Xi+\frac{\Xi^2}{4\lambda}-
\left(\frac{\Xi}{4\lambda}+\frac{H^2}2\right)\frac{\Box H}{v+H}.
\label{efflag}
\end{equation}
The advantage of writing the Lagrangian in this form is that the last term, still explicitly depending on the Higgs field [determined implicitly by Eq.~\eqref{higgs}], is suppressed by two extra derivatives as compared to the first two terms. If we are only interested in the dominant contributions to the effective Lagrangian for the NG field $U(\pi)$, containing the lowest number of derivatives, we can simply discard this last term and write
\begin{equation}
\La^{\text{LO}}_{\text{eff}}=v^2\Xi+\frac{\Xi^2}{4\lambda}.
\label{LO_eft}
\end{equation}
A similar expression can be derived for a more general class of Lagrangians,
\begin{equation}
\La=D_\mu\he\phi D^\mu\phi-V(\he\phi\phi),
\end{equation}
with a general potential $V$. While it does not seem feasible to obtain a closed expression analogous to Eq.~\eqref{efflag} in this general case, one can find an iterative solution of the equation of motion for the Higgs mode by a generalization of Eq.~\eqref{higgs}. Up to the second order in derivatives, the effective Lagrangian generalizing Eq.~\eqref{LO_eft} then reads
\begin{equation}
\La^{\text{LO}}_{\text{eff}}=v^2\Xi+\frac{\Xi^2}{2V''(v^2)};
\label{LO_eft_gen}
\end{equation}
the derivatives of the potential are with respect to $\he\phi\phi$, $v^2=\langle\he\phi\phi\rangle$ as before and $\Xi$ is still defined by Eq.~\eqref{Xidef}.

The effective Lagrangians~\eqref{LO_eft} and~\eqref{LO_eft_gen} determine the kinetic terms as well as the interactions of the NG bosons to the leading order in the derivative expansion. Note that in spite of presence of terms with a single time derivative, the Lagrangian is invariant under the internal $\gr{SU}(N)\times\gr{U(1)}$ symmetry. This is in accord with the general EFT for NG bosons in nonrelativistic systems\cite{Leutwyler1994a} and the fact that in classical relativistic scalar field theories with chemical potential, considered here, the ground state expectation value of all \emph{unbroken} symmetry generators is always zero.

%%%%%%%%%%%%%%%%%%%%%%%%%%%%%%%%%%%%%%%%%%%%%%%%%%%%%%%%%%%%%%%%%%%

\subsection{Example}
\label{subsec:example}

For illustration, let us show how to use the Lagrangian~\eqref{LO_eft} to determine the dispersion relations of the NG bosons in the simplest non-Abelian case with $N=2$.\cite{Miransky2002a,*Schaefer2001a} To that end, we only need to keep terms with up to two derivatives. Inserting the definition of $\Xi$ from Eq.~\eqref{Xidef}, the Lagrangian~\eqref{LO_eft} becomes
\begin{equation}
\begin{split}
\La^\text{LO}_\text{eff}={}&v^2\he\phi_0\de_\mu\he U\de^\mu U\phi_0+2\imag\mu v^2\he\phi_0\he U\de_0 U\phi_0\\
&-\frac{\mu^2}\lambda(\he\phi_0\he U\de_0U\phi_0)^2+\dotsb.
\end{split}
\label{kin}
\end{equation}
To proceed, we have to fix the orientation of the ground state and a parametrization of the matrix field $U(\pi)$ in terms of the three NG fields $\pi_{1,2,3}$ of the symmetry-breaking pattern $\gr{SU(2)\times U(1)}\to\gr{U(1)}'$. We set as usual $\phi_0=(0,1)^T$ and use the exponential parametrization
\begin{equation}
U(\pi)\equiv\exp\left(\frac\imag v\vec\pi\cdot\vec T\right),
\end{equation}
where $T_{1,2}\equiv\tau_{1,2}$ are two of the Pauli matrices and $T_3\equiv\openone$. The bilinear part of the Lagrangian~\eqref{kin} then becomes
\begin{equation}
\begin{split}
\La_{\text{bilin}}={}&(\de_\mu\pi_3)^2+\frac{\mu^2}{\lambda v^2}(\de_0\pi_3)^2\\
&+(\de_\mu\pi_{1,2})^2+2\mu(\pi_1\de_0\pi_2-\pi_2\de_0\pi_1)\\
={}&\left(1+\frac{\mu^2}{\lambda v^2}\right)(\de_0\pi_3)^2-(\vek\nabla\pi_3)^2\\
&+2\imag\mu\he\psi\de_0\psi+\de_\mu\he\psi\de^\mu\psi,
\label{kin2}
\end{split}
\end{equation}
where $\psi\equiv\pi_1-\imag\pi_2$. From here we conclude that $\pi_3$ annihilates a type-A NG boson\cite{Watanabe2012b,Watanabe2011a,*Hidaka2013b} with the squared phase velocity $c^2=(\mu^2-m^2)/(3\mu^2-m^2)$, while $\psi$ annihilates a type-B NG boson with the dispersion relation $E(\vek p)=\vek p^2/(2\mu)$ at low momenta such that $|\vek p|\ll\mu$, in agreement with Refs.~\onlinecite{Miransky2002a,*Schaefer2001a}. (See Refs.~\onlinecite{Brauner2010a} and~\onlinecite{Watanabe2020a,*Beekman2019a,*AlvarezGaume2020a} for a review of counting rules for NG bosons.) Note that the dispersion relation of the type-B NG mode is fixed by the $v^2\Xi$ term in the Lagrangian~\eqref{LO_eft}. On the other hand, to get the phase velocity of the type-A NG mode right, the $\Xi^2/(4\lambda)$ term in the Lagrangian~\eqref{LO_eft} is essential. It is clear from Eq.~\eqref{LO_eft_gen} that analogous results hold for more general scalar potentials than that in Eq.~\eqref{LagN}.

The above example demonstrates the utility of the EFT approach compared to the underlying linear sigma model. The dispersion relations of the NG modes can be extracted easily from the Lagrangian~\eqref{LO_eft} or~\eqref{LO_eft_gen} without having to deal with the mixing between the NG and Higgs fields.~\cite{Miransky2002a,*Schaefer2001a} The utility of the EFT would become even more apparent if we wanted to calculate scattering amplitudes of the NG modes. In this case, kinetic mixing would lead to substantial complications, including having to deal with matrix propagators and a nontrivial mapping between fields and the asymptotic one-particle states of the scattering process.\cite{Brauner2006b,*Brauner2018a}

Given the kinetic mixing between the NG and Higgs modes, implied by the microscopic Lagrangian~\eqref{LagN} for any nonzero $\mu$, one might wonder whether it is appropriate to integrate out the Higgs \emph{field} rather than the actual Higgs mode in the spectrum, as we did above. The answer is yes, and the reason why is illustrated in Appendix~\ref{app:Higgs} using a simple toy model.

Finally, let us remark on the connection of our effective Lagrangian~\eqref{LO_eft_gen} to the model-independent coset construction of effective Lagrangians for NG bosons.\cite{Coleman1969a,*Callan1969a} Equation~\eqref{Xidef} can be expressed in terms of the Lie-algebra-valued Maurer-Cartan (MC) form, whose gauged version is conventionally defined as
\begin{equation}
\omega_\mu\equiv-\imag\he UD_\mu U.
\end{equation}
Then $\Xi$ becomes
\begin{equation}
\Xi=\he\phi_0\omega_\mu\omega^\mu\phi_0-\mu^2.
\end{equation}
Sandwiching the square of the MC form between $\he\phi_0$ and $\phi_0$ projects out its components corresponding to spontaneously broken symmetry generators. The two terms in Eq.~\eqref{LO_eft_gen} thus naturally match invariants produced by the coset construction. The advantage of our procedure for obtaining the EFT explicitly from the microscopic Lagrangian is that it fixes the values of the low-energy couplings in the effective Lagrangian.

%%%%%%%%%%%%%%%%%%%%%%%%%%%%%%%%%%%%%%%%%%%%%%%%%%%%%%%%%%%%%%%%%%%

\section{Conclusions}
\label{sec:conclusions}

The present note addresses various aspects of description of finite-density systems using relativistic scalar field theory equipped with a chemical potential. The focus of the text is twofold.

First, in Sec.~\ref{sec:density}, we give an elementary but thorough discussion of the basic framework of classical field theory at nonzero density. Little of the presented material is truly new, yet we hope that the text sheds light on a number of subtle issues that are often glossed over. These include in particular (i) the precise correspondence of the two different realizations of nonzero particle density via time-dependent field configurations and via the addition of a chemical potential, and its violation in case of explicit symmetry breaking, (ii) the nature of breaking of Lorentz invariance by the chemical potential, and (iii) the identification of the Hamiltonian whose minimization yields the equilibrium many-body state.

Second, in Sec.~\ref{sec:EFT} we focused on the EFT description of NG bosons of global symmetries spontaneously broken by the dense medium. Starting from a scalar model, we showed how the EFT can be deduced explicitly by integrating out the massive (Higgs) degrees of freedom. This bridges the gap between the fully model-independent EFT description of NG bosons (which is universal but requires special techniques to realize the spontaneously broken symmetry nonlinearly\cite{Coleman1969a,*Callan1969a}), and simple linear sigma models (where symmetry is realized linearly, but general consequences of spontaneous symmetry breaking such as the very existence of NG bosons are not manifest).

The material of Sec.~\ref{sec:EFT} is of practical utility in that it allows one to study the physics of NG bosons below the mass scale of the Higgs mode(s) integrated out, without the need to deal with the mixing between the NG and Higgs degrees of freedom. One can contemplate generalization of the arguments in Sec.~\ref{sec:EFT} in several directions. While the discussion in Sec.~\ref{sec:EFT} was limited to an $N$-plet of complex fields in the fundamental representation of $\gr{SU}(N)$, one can analyze in a similar fashion models with a different symmetry group, or a different representation thereof. Likewise, one can include chemical potential(s) for other generators of the symmetry group. While an explicit expression for the effective Lagrangian applicable in such a general setting is not provided here, the EFT for NG bosons may be obtained case by case using the algorithm presented in Sec.~\ref{sec:EFT}.

%%%%%%%%%%%%%%%%%%%%%%%%%%%%%%%%%%%%%%%%%%%%%%%%%%%%%%%%%%%%%%%%%%%

\begin{acknowledgments}
I owe sincere thanks to Denis Parganlija and Andreas Schmitt for discussions that inspired large parts of Sec.~\ref{sec:density}, and to Hiroaki Abuki and Sergej Moroz for collaboration that motivated the ideas presented in Sec.~\ref{sec:EFT} and in the appendices. I would like to acknowledge financial support from the ToppForsk-UiS program of the University of Stavanger and the University Fund, Grant No.~PR-10614.
\end{acknowledgments}

%%%%%%%%%%%%%%%%%%%%%%%%%%%%%%%%%%%%%%%%%%%%%%%%%%%%%%%%%%%%%%%%%%%

\appendix

\section{Grand canonical Hamiltonian from generally covariant action}
\label{app:metric}

An established way to calculate the energy-momentum (EM) tensor of a given relativistic field theory is to couple it to background spacetime geometry and take a variation of the resulting generally covariant action with respect to the metric. In this appendix, we show that this approach is treacherous when applied to systems with nonzero chemical potential. For the sake of simplicity, we will assume that the symmetry-breaking parameter $j$ is zero throughout this appendix.

It is easy to promote the model~\eqref{Lag} to a generally covariant field theory,
\begin{equation}
S=\int\dd x\,\sqrt{-|g|}\,[g^{\mu\nu}D_\mu\phi^*D_\nu\phi-V(\phi^*\phi)].
\label{lsmcovariant}
\end{equation}
We have added the spacetime metric $g_{\mu\nu}(x)$ as well as a background gauge field $A_\mu(x)$, which enters the action through the covariant derivative, $D_\mu\phi\equiv(\de_\mu-\imag A_\mu)\phi$. The EM tensor can now be defined as
\begin{equation}
\Theta^{\mu\nu}=-\frac{2}{\sqrt{-|g|}}\frac{\delta S}{\delta g_{\mu\nu}}.
\label{Theta}
\end{equation}
Using the identities
\begin{equation}
\begin{split}
\delta g^{\mu\nu}&=-g^{\mu\alpha}g^{\nu\beta}\delta g_{\alpha\beta},\\
\delta\sqrt{-|g|}&=\frac12\sqrt{-|g|}\,g^{\mu\nu}\delta g_{\mu\nu},
\end{split}
\end{equation}
one finds the EM tensor for the theory of Eq.~\eqref{lsmcovariant},
\begin{equation}
\Theta^{\mu\nu}=(g^{\mu\alpha}g^{\nu\beta}+g^{\nu\alpha}g^{\mu\beta})D_\alpha\phi^*D_\beta\phi-g^{\mu\nu}\La.
\label{Thetamodel}
\end{equation}
We can now set the background spacetime geometry back to that of Minkowski spacetime and extract the Hamiltonian density,
\begin{equation}
\Ha=\Theta^{00}=D_0\phi^*D_0\phi+\vek\nabla\phi^*\cdot\vek\nabla\phi+V(\phi^*\phi).
\label{HaTheta}
\end{equation}
If we interpret $A_0$ as the chemical potential in line with our discussion in Sec.~\ref{sec:density}, we would expect Eq.~\eqref{HaTheta} to recover the grand canonical Hamiltonian~\eqref{muHamaux}. After all, the latter can certainly be obtained either by applying Noether's theorem to the grand canonical Lagrangian~\eqref{muLag}, or by the Legendre transform thereof. Quite surprisingly, Eq.~\eqref{HaTheta} does \emph{not} equal the grand canonical Hamiltonian, and as such cannot be used directly to study the many-body equilibrium state of our field theory.

Even more surprisingly, the problem lies in Eq.~\eqref{Theta} itself. This is not a definition, but rather a derived expression, which relies on the covariant transformation rules for all the fields in the theory under coordinate diffeomorphisms, in particular on
\begin{equation}
\delta\phi=-\xi^\mu\de_\mu\phi,\qquad
\delta A_\mu=-\xi^\nu\de_\nu A_\mu-A_\nu\de_\mu\xi^\nu
\end{equation}
under an infinitesimal coordinate shift $\delta x^\mu=\xi^\mu(x)$. Yet, when Noether's theorem is applied to the system in flat spacetime, a local translation is performed on all the fields by means of $\phi'(x')=\phi(x)$ and analogously for $A_\mu$. This means that if we want to derive the grand canonical Hamiltonian by the variation of a generally covariant action, we have to treat $A_\mu$ as a set of \emph{scalar} fields.\cite{Brauner2020a}

To that end, we trade the metric $g_{\mu\nu}$ for the covariant vielbein $e^A_\mu$ via
\begin{equation}
g_{\mu\nu}=\eta_{AB}e^A_\mu e^B_\nu,
\end{equation}
where $\eta_{AB}$ is the flat Minkowski spacetime metric. Covariant vector fields can then be turned into scalars using the dual vielbein $E^\mu_A$, satisfying $E^\mu_Be^A_\mu=\eta^A_B$. The covariant action Eq.~\eqref{lsmcovariant} thus needs to be replaced with
\begin{align}
\notag
S=\int\dd x\,|e|\,\Bigl[&\eta^{AB}(E^\mu_A\de_\mu\phi^*+\imag A_A\phi^*)(E^\nu_B\de_\nu\phi-\imag A_B\phi)\\
&-V(\phi^*\phi)\Bigr],
\end{align}
where $A_A\equiv E^\mu_AA_\mu$ is the scalarized background gauge field. Since the action now depends explicitly on the vielbein rather than the metric, a generalization of Eq.~\eqref{Theta} is needed,
\begin{equation}
T^\mu_{\phantom\mu\nu}=-\frac1{|e|}e^A_\nu\frac{\delta S}{\delta e^A_\mu}.
\label{TmuA}
\end{equation}
Using the identities
\begin{equation}
\delta E^\mu_A=-E^\nu_AE^\mu_B\delta e^B_\nu,\qquad
\delta|e|=|e|E^\mu_A\delta e^A_\mu,
\end{equation}
one arrives at the EM tensor
\begin{equation}
\begin{split}
T^{\mu\nu}={}&(g^{\mu\alpha}g^{\nu\beta}+g^{\nu\alpha}g^{\mu\beta})\de_\alpha\phi^*\de_\beta\phi-g^{\mu\nu}\La\\
&+\imag\eta^{AB}E^\mu_Ag^{\nu\alpha}A_B(\phi^*\de_\alpha\phi-\de_\alpha\phi^*\phi),
\end{split}
\label{Tmodel}
\end{equation}
to be compared to Eq.~\eqref{Thetamodel}. It is easy to check that, unlike the impostor Hamiltonian~\eqref{HaTheta}, the Hamiltonian obtained from $T^{00}$ by going to the flat-space limit \emph{does} agree with the grand canonical Hamiltonian~\eqref{muHamaux}.

One might still be puzzled over the fact that we found two manifestly different Hamiltonians just by using two different covariantizations of the same scalar field theory. This issue is closely related to the infamous discrepancy between the manifestly symmetric EM tensor~\eqref{Theta} and the ``canonical'' EM tensor obtained using Noether's theorem, in theories of higher-spin particles.\cite{Brauner2020a} The two EM tensors are generally equal up to terms that vanish upon using the equation of motion. Indeed, the EM tensors~\eqref{Thetamodel} and~\eqref{Tmodel} would equal, should the equation of motion for $A_\mu$ hold. The problem is that in our case, $A_\mu$ represents an external field that is \emph{not} subject to an equation of motion. The two EM tensors, as well as the corresponding Hamiltonians, are therefore genuinely physically different.

%%%%%%%%%%%%%%%%%%%%%%%%%%%%%%%%%%%%%%%%%%%%%%%%%%%%%%%%%%%%%%%%%%%

\section{Integrating out heavy modes}
\label{app:Higgs}

In Sec.~\ref{sec:EFT}, we obtained an EFT for the NG degrees of freedom by explicitly integrating out, at tree level, the Higgs degrees of freedom. We did so by eliminating the Higgs fields by using their equation of motion. However, the chemical potential as a rule induces mixing of Higgs and NG fields in the bilinear Lagrangian. As a result, the Higgs fields may couple to both NG and Higgs \emph{modes} in the spectrum. It is therefore natural to ask whether the procedure employed in Sec.~\ref{sec:EFT} is in fact correct, and to what extent it is unique.

To get an insight into this issue in as simple a setting as possible, we analyze in this appendix a simple toy model, defined by the Lagrangian
\begin{equation}
\La=\frac12(\de_\mu\phi_1)^2+\frac12(\de_\mu\phi_2)^2-\frac12m^2(\phi_1-\phi_2)^2,
\label{Lagapp}
\end{equation}
where $\phi_{1,2}$ are two real scalar fields and $m$ is a positive mass parameter. This is a noninteracting field theory, as is easily seen by defining $\phi_\pm\equiv(\phi_1\pm\phi_2)/\sqrt2$. The field $\phi_+$ annihilates a massless particle, namely a NG boson of the global symmetry of the Lagrangian~\eqref{Lagapp} under the shift  $\phi_{1,2}\to\phi_{1,2}+\theta$. The field $\phi_-$ annihilates a particle with the mass $m\sqrt2$. A low-energy EFT for the massless mode can be obtained by integrating out $\phi_-$, and it takes the expected form of a free massless scalar field theory.

Suppose, however, that we are not smart enough to notice the change of basis from $\phi_{1,2}$ to $\phi_\pm$. Let us instead try to derive a low-energy EFT for $\phi_1$ by integrating out $\phi_2$. The equation of motion for $\phi_2$, following from the Lagrangian~\eqref{Lagapp}, reads $\Box\phi_2+m^2(\phi_2-\phi_1)=0$, whence one obtains
\begin{equation}
\phi_2=m^2(\Box+m^2)^{-1}\phi_1.
\label{eom}
\end{equation}
Substituting this back into Eq.~\eqref{Lagapp}, one derives the effective Lagrangian for $\phi_1$ only,
\begin{equation}
\La_\text{eff}=-\frac12\phi_1\Box(\Box+2m^2)(\Box+m^2)^{-1}\phi_1.
\label{effLagapp}
\end{equation}
From this Lagrangian, we obtain in turn the propagator of $\phi_1$ in the effective theory,\footnote{Since the original theory~\eqref{Lagapp} is noninteracting, the propagator~\eqref{propapp} is the only nontrivial connected Green's function of the EFT.}
\begin{equation}
G^{-1}(p)=\frac{p^2-m^2}{p^2(p^2-2m^2)}.
\label{propapp}
\end{equation}
This is the same propagator of $\phi_1$ as we would find from the Lagrangian~\eqref{Lagapp}. This observation generalizes to interacting field theories as well: if we are able, at least in principle, to integrate out a set of fields, then the resulting, generally nonlocal EFT reproduces exactly all the Green's functions of the remaining fields.

A more subtle question is to what extent Eq.~\eqref{effLagapp} is a~\emph{low-energy} EFT to Eq.~\eqref{Lagapp}. We have not really integrated out the heavy mode $\phi_-$. Indeed, both fields $\phi_{1,2}$ contain admixtures of both $\phi_\pm$ and therefore couple to both states in the spectrum. We could have instead chosen to integrate out $\phi_1$ and would thus have arrived at an equivalent effective Lagrangian for $\phi_2$. Accordingly, the propagator~\eqref{propapp} obtained
from Eq.~\eqref{effLagapp} still features two poles, at $p^2=0$ and $p^2=2m^2$.

The key to understanding this issue lies in the observation that off-shell Green's functions depend on the precise definition of the fields in the theory. The detailed form of the effective Lagrangian may therefore depend on what field we choose to integrate out. It is the on-shell Green's functions that must be correctly reproduced by the low-energy EFT. In our simple noninteracting model, this reduces to the poles in the propagator. Once we are only interested in the low-energy physics of the NG modes, we only need to keep the contribution of the massless poles to the Green's functions; any massive pole corresponding to a particle of mass $m$ can be expanded in powers of $p^2/m^2$. In terms of Eq.~\eqref{effLagapp} this means that the low-energy EFT for the NG mode is obtained by expanding the Lagrangian in powers of $\Box/m^2$. The expanded Lagrangian contains an infinite number of terms, but it describes the same physics as the effective Lagrangian for $\phi_+$: a single massless noninteracting scalar field.

While this example is rather trivial, the same logic applies to the derivation of the low-energy EFT for NG modes in Sec.~\ref{sec:EFT}. It may seem that due to the mixing in the kinetic term, integrating out the Higgs fields is not the same as integrating out the Higgs modes from the spectrum. However, the resulting EFT correctly reproduces the full (off-shell) Green's functions of the NG fields. By performing the expansion of the effective Lagrangian in powers of derivatives of the NG fields, one in effect only keeps the part of the Green's functions containing the NG poles. The expanded effective Lagrangian then correctly reproduces the on-shell scattering amplitudes of the NG bosons. The above argument relies only on the analytic structure of the Green's functions and not on the particular way of integrating out the heavy fields. It therefore applies equally well to the full quantum effective action of the theory as well as its tree-level approximation, obtained by using the equation of motion of the heavy fields.

%%%%%%%%%%%%%%%%%%%%%%%%%%%%%%%%%%%%%%%%%%%%%%%%%%%%%%%%%%%%%%%%%%%

\bibliography{references}

\begin{thebibliography}{36}%
\makeatletter
\providecommand \@ifxundefined [1]{%
 \@ifx{#1\undefined}
}%
\providecommand \@ifnum [1]{%
 \ifnum #1\expandafter \@firstoftwo
 \else \expandafter \@secondoftwo
 \fi
}%
\providecommand \@ifx [1]{%
 \ifx #1\expandafter \@firstoftwo
 \else \expandafter \@secondoftwo
 \fi
}%
\providecommand \natexlab [1]{#1}%
\providecommand \enquote  [1]{``#1''}%
\providecommand \bibnamefont  [1]{#1}%
\providecommand \bibfnamefont [1]{#1}%
\providecommand \citenamefont [1]{#1}%
\providecommand \href@noop [0]{\@secondoftwo}%
\providecommand \href [0]{\begingroup \@sanitize@url \@href}%
\providecommand \@href[1]{\@@startlink{#1}\@@href}%
\providecommand \@@href[1]{\endgroup#1\@@endlink}%
\providecommand \@sanitize@url [0]{\catcode `\\12\catcode `\$12\catcode
  `\&12\catcode `\#12\catcode `\^12\catcode `\_12\catcode `\%12\relax}%
\providecommand \@@startlink[1]{}%
\providecommand \@@endlink[0]{}%
\providecommand \url  [0]{\begingroup\@sanitize@url \@url }%
\providecommand \@url [1]{\endgroup\@href {#1}{\urlprefix }}%
\providecommand \urlprefix  [0]{URL }%
\providecommand \Eprint [0]{\href }%
\providecommand \doibase [0]{http://dx.doi.org/}%
\providecommand \selectlanguage [0]{\@gobble}%
\providecommand \bibinfo  [0]{\@secondoftwo}%
\providecommand \bibfield  [0]{\@secondoftwo}%
\providecommand \translation [1]{[#1]}%
\providecommand \BibitemOpen [0]{}%
\providecommand \bibitemStop [0]{}%
\providecommand \bibitemNoStop [0]{.\EOS\space}%
\providecommand \EOS [0]{\spacefactor3000\relax}%
\providecommand \BibitemShut  [1]{\csname bibitem#1\endcsname}%
\let\auto@bib@innerbib\@empty
%</preamble>
\bibitem [{\citenamefont {Ginzburg}\ and\ \citenamefont
  {Landau}(1950)}]{Ginzburg1950a}%
  \BibitemOpen
  \bibfield  {author} {\bibinfo {author} {\bibfnamefont {V.~L.}\ \bibnamefont
  {Ginzburg}}\ and\ \bibinfo {author} {\bibfnamefont {L.~D.}\ \bibnamefont
  {Landau}},\ }\href@noop {} {\bibfield  {journal} {\bibinfo  {journal} {Zh.
  Eksp. Teor. Fiz.}\ }\textbf {\bibinfo {volume} {20}},\ \bibinfo {pages}
  {1064} (\bibinfo {year} {1950})}\BibitemShut {NoStop}%
\bibitem [{\citenamefont {Goldstone}(1961)}]{Goldstone1961a}%
  \BibitemOpen
  \bibfield  {author} {\bibinfo {author} {\bibfnamefont {J.}~\bibnamefont
  {Goldstone}},\ }\href {\doibase 10.1007/BF02812722} {\bibfield  {journal}
  {\bibinfo  {journal} {Nuovo Cim.}\ }\textbf {\bibinfo {volume} {19}},\
  \bibinfo {pages} {154} (\bibinfo {year} {1961})}\BibitemShut {NoStop}%
%%CITATION = NUCIA,19,154;%%
\bibitem [{\citenamefont {Goldstone}\ \emph {et~al.}(1962)\citenamefont
  {Goldstone}, \citenamefont {Salam},\ and\ \citenamefont
  {Weinberg}}]{Goldstone1962a}%
  \BibitemOpen
  \bibfield  {author} {\bibinfo {author} {\bibfnamefont {J.}~\bibnamefont
  {Goldstone}}, \bibinfo {author} {\bibfnamefont {A.}~\bibnamefont {Salam}}, \
  and\ \bibinfo {author} {\bibfnamefont {S.}~\bibnamefont {Weinberg}},\ }\href
  {\doibase 10.1103/PhysRev.127.965} {\bibfield  {journal} {\bibinfo  {journal}
  {Phys. Rev.}\ }\textbf {\bibinfo {volume} {127}},\ \bibinfo {pages} {965}
  (\bibinfo {year} {1962})}\BibitemShut {NoStop}%
%%CITATION = PHRVA,127,965;%%
\bibitem [{\citenamefont {Higgs}(1964)}]{Higgs1964a}%
  \BibitemOpen
  \bibfield  {author} {\bibinfo {author} {\bibfnamefont {P.~W.}\ \bibnamefont
  {Higgs}},\ }\href {\doibase 10.1103/PhysRevLett.13.508} {\bibfield  {journal}
  {\bibinfo  {journal} {Phys. Rev. Lett.}\ }\textbf {\bibinfo {volume} {13}},\
  \bibinfo {pages} {508} (\bibinfo {year} {1964})}\BibitemShut {NoStop}%
\bibitem [{\citenamefont {Nielsen}\ and\ \citenamefont
  {Chadha}(1976)}]{Nielsen1976a}%
  \BibitemOpen
  \bibfield  {author} {\bibinfo {author} {\bibfnamefont {H.~B.}\ \bibnamefont
  {Nielsen}}\ and\ \bibinfo {author} {\bibfnamefont {S.}~\bibnamefont
  {Chadha}},\ }\href {\doibase 10.1016/0550-3213(76)90025-0} {\bibfield
  {journal} {\bibinfo  {journal} {Nucl. Phys.}\ }\textbf {\bibinfo {volume}
  {B105}},\ \bibinfo {pages} {445} (\bibinfo {year} {1976})}\BibitemShut
  {NoStop}%
%%CITATION = NUPHA,B105,445;%%
\bibitem [{\citenamefont {Nambu}(2004)}]{Nambu2004a}%
  \BibitemOpen
  \bibfield  {author} {\bibinfo {author} {\bibfnamefont {Y.}~\bibnamefont
  {Nambu}},\ }\href {\doibase 10.1023/B:JOSS.0000019827.74407.2d} {\bibfield
  {journal} {\bibinfo  {journal} {J. Stat. Phys.}\ }\textbf {\bibinfo {volume}
  {115}},\ \bibinfo {pages} {7} (\bibinfo {year} {2004})}\BibitemShut {NoStop}%
%%CITATION = JSTPB,115,7;%%
\bibitem [{\citenamefont {Leutwyler}(1994{\natexlab{a}})}]{Leutwyler1994a}%
  \BibitemOpen
  \bibfield  {author} {\bibinfo {author} {\bibfnamefont {H.}~\bibnamefont
  {Leutwyler}},\ }\href {\doibase 10.1103/PhysRevD.49.3033} {\bibfield
  {journal} {\bibinfo  {journal} {Phys. Rev.}\ }\textbf {\bibinfo {volume}
  {D49}},\ \bibinfo {pages} {3033} (\bibinfo {year} {1994}{\natexlab{a}})},\
  \Eprint {http://arxiv.org/abs/hep-ph/9311264} {arXiv:hep-ph/9311264}
  \BibitemShut {NoStop}%
%%CITATION = HEP-PH/9311264;%%
\bibitem [{\citenamefont {Miransky}\ and\ \citenamefont
  {Shovkovy}(2002)}]{Miransky2002a}%
  \BibitemOpen
  \bibfield  {author} {\bibinfo {author} {\bibfnamefont {V.~A.}\ \bibnamefont
  {Miransky}}\ and\ \bibinfo {author} {\bibfnamefont {I.~A.}\ \bibnamefont
  {Shovkovy}},\ }\href {\doibase 10.1103/PhysRevLett.88.111601} {\bibfield
  {journal} {\bibinfo  {journal} {Phys. Rev. Lett.}\ }\textbf {\bibinfo
  {volume} {88}},\ \bibinfo {pages} {111601} (\bibinfo {year} {2002})},\
  \Eprint {http://arxiv.org/abs/hep-ph/0108178} {arXiv:hep-ph/0108178}
  \BibitemShut {NoStop}%
%%CITATION = HEP-PH/0108178;%%
\bibitem [{\citenamefont {{Sch\"afer}}\ \emph {et~al.}(2001)\citenamefont
  {{Sch\"afer}}, \citenamefont {Son}, \citenamefont {Stephanov}, \citenamefont
  {Toublan},\ and\ \citenamefont {Verbaarschot}}]{Schaefer2001a}%
  \BibitemOpen
  \bibfield  {author} {\bibinfo {author} {\bibfnamefont {T.}~\bibnamefont
  {{Sch\"afer}}}, \bibinfo {author} {\bibfnamefont {D.~T.}\ \bibnamefont
  {Son}}, \bibinfo {author} {\bibfnamefont {M.~A.}\ \bibnamefont {Stephanov}},
  \bibinfo {author} {\bibfnamefont {D.}~\bibnamefont {Toublan}}, \ and\
  \bibinfo {author} {\bibfnamefont {J.~J.~M.}\ \bibnamefont {Verbaarschot}},\
  }\href {\doibase 10.1016/S0370-2693(01)01265-5} {\bibfield  {journal}
  {\bibinfo  {journal} {Phys. Lett.}\ }\textbf {\bibinfo {volume} {B522}},\
  \bibinfo {pages} {67} (\bibinfo {year} {2001})},\ \Eprint
  {http://arxiv.org/abs/hep-ph/0108210} {arXiv:hep-ph/0108210} \BibitemShut
  {NoStop}%
%%CITATION = HEP-PH/0108210;%%
\bibitem [{\citenamefont {Brauner}(2005)}]{Brauner2005a}%
  \BibitemOpen
  \bibfield  {author} {\bibinfo {author} {\bibfnamefont {T.}~\bibnamefont
  {Brauner}},\ }\href {\doibase 10.1103/PhysRevD.72.076002} {\bibfield
  {journal} {\bibinfo  {journal} {Phys. Rev.}\ }\textbf {\bibinfo {volume}
  {D72}},\ \bibinfo {pages} {076002} (\bibinfo {year} {2005})},\ \Eprint
  {http://arxiv.org/abs/hep-ph/0508011} {arXiv:hep-ph/0508011} \BibitemShut
  {NoStop}%
%%CITATION = HEP-PH/0508011;%%
\bibitem [{\citenamefont {Brauner}(2010)}]{Brauner2010a}%
  \BibitemOpen
  \bibfield  {author} {\bibinfo {author} {\bibfnamefont {T.}~\bibnamefont
  {Brauner}},\ }\href {\doibase 10.3390/sym2020609} {\bibfield  {journal}
  {\bibinfo  {journal} {Symmetry}\ }\textbf {\bibinfo {volume} {2}},\ \bibinfo
  {pages} {609} (\bibinfo {year} {2010})},\ \Eprint
  {http://arxiv.org/abs/1001.5212} {arXiv:1001.5212 [hep-th]} \BibitemShut
  {NoStop}%
%%CITATION = ARXIV:1001.5212;%%
\bibitem [{\citenamefont {Watanabe}\ and\ \citenamefont
  {Murayama}(2012)}]{Watanabe2012b}%
  \BibitemOpen
  \bibfield  {author} {\bibinfo {author} {\bibfnamefont {H.}~\bibnamefont
  {Watanabe}}\ and\ \bibinfo {author} {\bibfnamefont {H.}~\bibnamefont
  {Murayama}},\ }\href {\doibase 10.1103/PhysRevLett.108.251602} {\bibfield
  {journal} {\bibinfo  {journal} {Phys. Rev. Lett.}\ }\textbf {\bibinfo
  {volume} {108}},\ \bibinfo {pages} {251602} (\bibinfo {year} {2012})},\
  \Eprint {http://arxiv.org/abs/1203.0609} {arXiv:1203.0609 [hep-th]}
  \BibitemShut {NoStop}%
%%CITATION = ARXIV:1203.0609;%%
\bibitem [{\citenamefont {Watanabe}\ and\ \citenamefont
  {Murayama}(2014)}]{Watanabe2014a}%
  \BibitemOpen
  \bibfield  {author} {\bibinfo {author} {\bibfnamefont {H.}~\bibnamefont
  {Watanabe}}\ and\ \bibinfo {author} {\bibfnamefont {H.}~\bibnamefont
  {Murayama}},\ }\href {\doibase 10.1103/PhysRevX.4.031057} {\bibfield
  {journal} {\bibinfo  {journal} {Phys. Rev.}\ }\textbf {\bibinfo {volume}
  {X4}},\ \bibinfo {pages} {031057} (\bibinfo {year} {2014})},\ \Eprint
  {http://arxiv.org/abs/1402.7066} {arXiv:1402.7066 [hep-th]} \BibitemShut
  {NoStop}%
%%CITATION = ARXIV:1402.7066;%%
\bibitem [{Note1()}]{Note1}%
  \BibitemOpen
  \bibinfo {note} {The Hamiltonian $\protect \mathscr {H}_\mu $ can be
  minimized in the same way even for nonzero $j$, but then the time evolution
  implied by the relation $\Pi =\protect \mathaccentV {dot}05F\phi ^*$ takes
  the field outside the thus found ground state manifold.}\BibitemShut {Stop}%
\bibitem [{\citenamefont {Nicolis}\ and\ \citenamefont
  {Piazza}(2013)}]{Nicolis2013a}%
  \BibitemOpen
  \bibfield  {author} {\bibinfo {author} {\bibfnamefont {A.}~\bibnamefont
  {Nicolis}}\ and\ \bibinfo {author} {\bibfnamefont {F.}~\bibnamefont
  {Piazza}},\ }\href {\doibase 10.1103/PhysRevLett.110.011602} {\bibfield
  {journal} {\bibinfo  {journal} {Phys. Rev. Lett.}\ }\textbf {\bibinfo
  {volume} {110}},\ \bibinfo {pages} {011602} (\bibinfo {year} {2013})},\
  \Eprint {http://arxiv.org/abs/1204.1570} {arXiv:1204.1570 [hep-th]}
  \BibitemShut {NoStop}%
%%CITATION = ARXIV:1204.1570;%%
\bibitem [{\citenamefont {Watanabe}\ \emph {et~al.}(2013)\citenamefont
  {Watanabe}, \citenamefont {Brauner},\ and\ \citenamefont
  {Murayama}}]{Watanabe2013b}%
  \BibitemOpen
  \bibfield  {author} {\bibinfo {author} {\bibfnamefont {H.}~\bibnamefont
  {Watanabe}}, \bibinfo {author} {\bibfnamefont {T.}~\bibnamefont {Brauner}}, \
  and\ \bibinfo {author} {\bibfnamefont {H.}~\bibnamefont {Murayama}},\ }\href
  {\doibase 10.1103/PhysRevLett.111.021601} {\bibfield  {journal} {\bibinfo
  {journal} {Phys. Rev. Lett.}\ }\textbf {\bibinfo {volume} {111}},\ \bibinfo
  {pages} {021601} (\bibinfo {year} {2013})},\ \Eprint
  {http://arxiv.org/abs/1303.1527} {arXiv:1303.1527 [hep-th]} \BibitemShut
  {NoStop}%
%%CITATION = ARXIV:1303.1527;%%
\bibitem [{\citenamefont {Kapusta}(1981)}]{Kapusta1981a}%
  \BibitemOpen
  \bibfield  {author} {\bibinfo {author} {\bibfnamefont {J.~I.}\ \bibnamefont
  {Kapusta}},\ }\href {\doibase 10.1103/PhysRevD.24.426} {\bibfield  {journal}
  {\bibinfo  {journal} {Phys. Rev.}\ }\textbf {\bibinfo {volume} {D24}},\
  \bibinfo {pages} {426} (\bibinfo {year} {1981})}\BibitemShut {NoStop}%
%%CITATION = PHRVA,D24,426;%%
\bibitem [{Note2()}]{Note2}%
  \BibitemOpen
  \bibinfo {note} {It is commonplace to assume that a given low-energy EFT can
  be ``UV-completed'' to a microscopic renormalizable field theory, although
  there are other logical possibilities. On the one hand, the microscopic
  theory need not be a local field theory. On the other hand, it is possible
  that there is no complete microscopic theory at all, and one instead finds an
  infinite tower of EFTs with successively broader range of validity.\cite
  {Castellani2002a}}\BibitemShut {NoStop}%
\bibitem [{\citenamefont {Leutwyler}(1994{\natexlab{b}})}]{Leutwyler1994b}%
  \BibitemOpen
  \bibfield  {author} {\bibinfo {author} {\bibfnamefont {H.}~\bibnamefont
  {Leutwyler}},\ }\href {\doibase 10.1006/aphy.1994.1094} {\bibfield  {journal}
  {\bibinfo  {journal} {Ann. Phys.}\ }\textbf {\bibinfo {volume} {235}},\
  \bibinfo {pages} {165} (\bibinfo {year} {1994}{\natexlab{b}})},\ \Eprint
  {http://arxiv.org/abs/hep-ph/9311274} {arXiv:hep-ph/9311274} \BibitemShut
  {NoStop}%
%%CITATION = HEP-PH/9311274;%%
\bibitem [{\citenamefont {Son}(2002)}]{Son2002a}%
  \BibitemOpen
  \bibfield  {author} {\bibinfo {author} {\bibfnamefont {D.~T.}\ \bibnamefont
  {Son}},\ }\href@noop {} {\  (\bibinfo {year} {2002})},\ \bibinfo {note}
  {arXiv:hep-ph/0204199}\BibitemShut {NoStop}%
%%CITATION = HEP-PH/0204199;%%
\bibitem [{\citenamefont {Brauner}(2020)}]{Brauner2020a}%
  \BibitemOpen
  \bibfield  {author} {\bibinfo {author} {\bibfnamefont {T.}~\bibnamefont
  {Brauner}},\ }\href {\doibase 10.1088/1402-4896/ab50a5} {\bibfield  {journal}
  {\bibinfo  {journal} {Phys. Scr.}\ }\textbf {\bibinfo {volume} {95}},\
  \bibinfo {pages} {035004} (\bibinfo {year} {2020})},\ \Eprint
  {http://arxiv.org/abs/1910.12224} {arXiv:1910.12224 [hep-th]} \BibitemShut
  {NoStop}%
\bibitem [{\citenamefont {Nicolis}\ \emph {et~al.}(2013)\citenamefont
  {Nicolis}, \citenamefont {Penco}, \citenamefont {Piazza},\ and\ \citenamefont
  {Rosen}}]{Nicolis2013b}%
  \BibitemOpen
  \bibfield  {author} {\bibinfo {author} {\bibfnamefont {A.}~\bibnamefont
  {Nicolis}}, \bibinfo {author} {\bibfnamefont {R.}~\bibnamefont {Penco}},
  \bibinfo {author} {\bibfnamefont {F.}~\bibnamefont {Piazza}}, \ and\ \bibinfo
  {author} {\bibfnamefont {R.~A.}\ \bibnamefont {Rosen}},\ }\href {\doibase
  10.1007/JHEP11(2013)055} {\bibfield  {journal} {\bibinfo  {journal} {JHEP}\
  }\textbf {\bibinfo {volume} {11}},\ \bibinfo {pages} {055} (\bibinfo {year}
  {2013})},\ \Eprint {http://arxiv.org/abs/1306.1240} {arXiv:1306.1240
  [hep-th]} \BibitemShut {NoStop}%
%%CITATION = ARXIV:1306.1240;%%
\bibitem [{\citenamefont {Nicolis}\ \emph {et~al.}(2015)\citenamefont
  {Nicolis}, \citenamefont {Penco}, \citenamefont {Piazza},\ and\ \citenamefont
  {Rattazzi}}]{Nicolis2015a}%
  \BibitemOpen
  \bibfield  {author} {\bibinfo {author} {\bibfnamefont {A.}~\bibnamefont
  {Nicolis}}, \bibinfo {author} {\bibfnamefont {R.}~\bibnamefont {Penco}},
  \bibinfo {author} {\bibfnamefont {F.}~\bibnamefont {Piazza}}, \ and\ \bibinfo
  {author} {\bibfnamefont {R.}~\bibnamefont {Rattazzi}},\ }\href {\doibase
  10.1007/JHEP06(2015)155} {\bibfield  {journal} {\bibinfo  {journal} {JHEP}\
  }\textbf {\bibinfo {volume} {06}},\ \bibinfo {pages} {155} (\bibinfo {year}
  {2015})},\ \Eprint {http://arxiv.org/abs/1501.03845} {arXiv:1501.03845
  [hep-th]} \BibitemShut {NoStop}%
%%CITATION = ARXIV:1501.03845;%%
\bibitem [{\citenamefont {Andersen}(2007)}]{Andersen2007a}%
  \BibitemOpen
  \bibfield  {author} {\bibinfo {author} {\bibfnamefont {J.~O.}\ \bibnamefont
  {Andersen}},\ }\href {\doibase 10.1103/PhysRevD.75.065011} {\bibfield
  {journal} {\bibinfo  {journal} {Phys. Rev.}\ }\textbf {\bibinfo {volume}
  {D75}},\ \bibinfo {pages} {065011} (\bibinfo {year} {2007})},\ \Eprint
  {http://arxiv.org/abs/hep-ph/0609020} {arXiv:hep-ph/0609020} \BibitemShut
  {NoStop}%
%%CITATION = HEP-PH/0609020;%%
\bibitem [{Note3()}]{Note3}%
  \BibitemOpen
  \bibinfo {note} {For positive $m^2$, a chemical potential $\mu >m$ is
  required; this is relativistic Bose-Einstein condensation. For negative
  $m^2$, the ground state is nontrivial even for zero chemical potential; this
  is the usual Higgs model.}\BibitemShut {Stop}%
\bibitem [{\citenamefont {Watanabe}\ and\ \citenamefont
  {Brauner}(2011)}]{Watanabe2011a}%
  \BibitemOpen
  \bibfield  {author} {\bibinfo {author} {\bibfnamefont {H.}~\bibnamefont
  {Watanabe}}\ and\ \bibinfo {author} {\bibfnamefont {T.}~\bibnamefont
  {Brauner}},\ }\href {\doibase 10.1103/PhysRevD.84.125013} {\bibfield
  {journal} {\bibinfo  {journal} {Phys. Rev.}\ }\textbf {\bibinfo {volume}
  {D84}},\ \bibinfo {pages} {125013} (\bibinfo {year} {2011})},\ \Eprint
  {http://arxiv.org/abs/1109.6327} {arXiv:1109.6327 [hep-ph]} \BibitemShut
  {NoStop}%
%%CITATION = ARXIV:1109.6327;%%
\bibitem [{\citenamefont {Hidaka}(2013)}]{Hidaka2013b}%
  \BibitemOpen
  \bibfield  {author} {\bibinfo {author} {\bibfnamefont {Y.}~\bibnamefont
  {Hidaka}},\ }\href {\doibase 10.1103/PhysRevLett.110.091601} {\bibfield
  {journal} {\bibinfo  {journal} {Phys. Rev. Lett.}\ }\textbf {\bibinfo
  {volume} {110}},\ \bibinfo {pages} {091601} (\bibinfo {year} {2013})},\
  \Eprint {http://arxiv.org/abs/1203.1494} {arXiv:1203.1494 [hep-th]}
  \BibitemShut {NoStop}%
%%CITATION = ARXIV:1203.1494;%%
\bibitem [{\citenamefont {Watanabe}(2020)}]{Watanabe2020a}%
  \BibitemOpen
  \bibfield  {author} {\bibinfo {author} {\bibfnamefont {H.}~\bibnamefont
  {Watanabe}},\ }\href {\doibase 10.1146/annurev-conmatphys-031119-050644}
  {\bibfield  {journal} {\bibinfo  {journal} {Ann. Rev. Condensed Matter
  Phys.}\ }\textbf {\bibinfo {volume} {11}},\ \bibinfo {pages} {169} (\bibinfo
  {year} {2020})},\ \Eprint {http://arxiv.org/abs/1904.00569} {arXiv:1904.00569
  [cond-mat.other]} \BibitemShut {NoStop}%
\bibitem [{\citenamefont {Beekman}\ \emph {et~al.}(2019)\citenamefont
  {Beekman}, \citenamefont {Rademaker},\ and\ \citenamefont {van
  Wezel}}]{Beekman2019a}%
  \BibitemOpen
  \bibfield  {author} {\bibinfo {author} {\bibfnamefont {A.~J.}\ \bibnamefont
  {Beekman}}, \bibinfo {author} {\bibfnamefont {L.}~\bibnamefont {Rademaker}},
  \ and\ \bibinfo {author} {\bibfnamefont {J.}~\bibnamefont {van Wezel}},\
  }\href {\doibase 10.21468/SciPostPhysLectNotes.11} {\bibfield  {journal}
  {\bibinfo  {journal} {SciPost Phys. Lect. Notes}\ }\textbf {\bibinfo {volume}
  {11}} (\bibinfo {year} {2019}),\ 10.21468/SciPostPhysLectNotes.11},\ \Eprint
  {http://arxiv.org/abs/1909.01820} {arXiv:1909.01820 [hep-th]} \BibitemShut
  {NoStop}%
\bibitem [{\citenamefont {\'Alvarez-Gaum\'e}\ \emph {et~al.}(2020)\citenamefont
  {\'Alvarez-Gaum\'e}, \citenamefont {Orlando},\ and\ \citenamefont
  {Reffert}}]{AlvarezGaume2020a}%
  \BibitemOpen
  \bibfield  {author} {\bibinfo {author} {\bibfnamefont {L.}~\bibnamefont
  {\'Alvarez-Gaum\'e}}, \bibinfo {author} {\bibfnamefont {D.}~\bibnamefont
  {Orlando}}, \ and\ \bibinfo {author} {\bibfnamefont {S.}~\bibnamefont
  {Reffert}},\ }\href@noop {} {\  (\bibinfo {year} {2020})},\ \Eprint
  {http://arxiv.org/abs/2008.03308} {arXiv:2008.03308 [hep-th]} \BibitemShut
  {NoStop}%
\bibitem [{\citenamefont {Brauner}(2006)}]{Brauner2006b}%
  \BibitemOpen
  \bibfield  {author} {\bibinfo {author} {\bibfnamefont {T.}~\bibnamefont
  {Brauner}},\ }\href {\doibase 10.1103/PhysRevD.74.085010} {\bibfield
  {journal} {\bibinfo  {journal} {Phys. Rev.}\ }\textbf {\bibinfo {volume}
  {D74}},\ \bibinfo {pages} {085010} (\bibinfo {year} {2006})},\ \Eprint
  {http://arxiv.org/abs/hep-ph/0607102} {arXiv:hep-ph/0607102} \BibitemShut
  {NoStop}%
%%CITATION = HEP-PH/0607102;%%
\bibitem [{\citenamefont {Brauner}\ and\ \citenamefont
  {Jakobsen}(2018)}]{Brauner2018a}%
  \BibitemOpen
  \bibfield  {author} {\bibinfo {author} {\bibfnamefont {T.}~\bibnamefont
  {Brauner}}\ and\ \bibinfo {author} {\bibfnamefont {M.~F.}\ \bibnamefont
  {Jakobsen}},\ }\href {\doibase 10.1103/PhysRevD.97.025021} {\bibfield
  {journal} {\bibinfo  {journal} {Phys. Rev.}\ }\textbf {\bibinfo {volume}
  {D97}},\ \bibinfo {pages} {025021} (\bibinfo {year} {2018})},\ \Eprint
  {http://arxiv.org/abs/1709.01251} {arXiv:1709.01251 [hep-th]} \BibitemShut
  {NoStop}%
%%CITATION = ARXIV:1709.01251;%%
\bibitem [{\citenamefont {Coleman}\ \emph {et~al.}(1969)\citenamefont
  {Coleman}, \citenamefont {Wess},\ and\ \citenamefont
  {Zumino}}]{Coleman1969a}%
  \BibitemOpen
  \bibfield  {author} {\bibinfo {author} {\bibfnamefont {S.~R.}\ \bibnamefont
  {Coleman}}, \bibinfo {author} {\bibfnamefont {J.}~\bibnamefont {Wess}}, \
  and\ \bibinfo {author} {\bibfnamefont {B.}~\bibnamefont {Zumino}},\ }\href
  {\doibase 10.1103/PhysRev.177.2239} {\bibfield  {journal} {\bibinfo
  {journal} {Phys. Rev.}\ }\textbf {\bibinfo {volume} {177}},\ \bibinfo {pages}
  {2239} (\bibinfo {year} {1969})}\BibitemShut {NoStop}%
%%CITATION = PHRVA,177,2239;%%
\bibitem [{\citenamefont {Callan}\ \emph {et~al.}(1969)\citenamefont {Callan},
  \citenamefont {Coleman}, \citenamefont {Wess},\ and\ \citenamefont
  {Zumino}}]{Callan1969a}%
  \BibitemOpen
  \bibfield  {author} {\bibinfo {author} {\bibfnamefont {C.~G.}\ \bibnamefont
  {Callan}}, \bibinfo {author} {\bibfnamefont {S.}~\bibnamefont {Coleman}},
  \bibinfo {author} {\bibfnamefont {J.}~\bibnamefont {Wess}}, \ and\ \bibinfo
  {author} {\bibfnamefont {B.}~\bibnamefont {Zumino}},\ }\href {\doibase
  10.1103/PhysRev.177.2247} {\bibfield  {journal} {\bibinfo  {journal} {Phys.
  Rev.}\ }\textbf {\bibinfo {volume} {177}},\ \bibinfo {pages} {2247} (\bibinfo
  {year} {1969})}\BibitemShut {NoStop}%
%%CITATION = PHRVA,177,2247;%%
\bibitem [{Note4()}]{Note4}%
  \BibitemOpen
  \bibinfo {note} {Since the original theory~\protect \textup {\hbox
  {\mathsurround \z@ \protect \normalfont (\ignorespaces \ref {Lagapp}\unskip
  \@@italiccorr )}} is noninteracting, the propagator~\protect \textup {\hbox
  {\mathsurround \z@ \protect \normalfont (\ignorespaces \ref {propapp}\unskip
  \@@italiccorr )}} is the only nontrivial connected Green's function of the
  EFT.}\BibitemShut {Stop}%
\bibitem [{\citenamefont {Castellani}(2002)}]{Castellani2002a}%
  \BibitemOpen
  \bibfield  {author} {\bibinfo {author} {\bibfnamefont {E.}~\bibnamefont
  {Castellani}},\ }\href {\doibase 10.1016/S1355-2198(02)00003-5} {\bibfield
  {journal} {\bibinfo  {journal} {Stud. Hist. Phil. Sci. B}\ }\textbf {\bibinfo
  {volume} {33}},\ \bibinfo {pages} {251} (\bibinfo {year} {2002})},\ \Eprint
  {http://arxiv.org/abs/physics/0101039} {arXiv:physics/0101039} \BibitemShut
  {NoStop}%
\end{thebibliography}%

\end{document}